\documentclass[11pt]{article}%{smfart}%
\usepackage{amssymb}\usepackage{amsmath} 
\newcommand{\fr}[2]{{\textstyle \frac{#1}{#2} }}
\newcommand{\nn}{\nonumber}
\newcommand{\tr}{\mathop{\rm tr}\nolimits}

\newcommand{\im}{{\mathbb I}}
\def\ve{\varepsilon}

\def\ot{\otimes}

\def\qed{\hfill $\square$}

\def\la{\lambda}
\begin{document} 
\hfill {\small 11/2011  }
\begin{center}

{\Large\bf On quantum L--operator \\[1mm]
 for two--dimensional lattice Toda model} \\[5mm]
{\sc Andrei Bytsko,${}^{1,2}$\hspace{8mm} 
Irina Davydenkova$\,{}^{2}$ } \\ [4mm]
{ \small
$^{1}$  Steklov Mathematics Institute\\
 Fontanka 27, 191023, St.~Petersburg, Russia  \\ [2mm]
$^{2}$ Section of Mathematics, University of Geneva\\ 
 2--4 rue de Li\`evre, C.P. 64, 1211 Gen\`eve 4, Switzerland  
} \\ [2.5mm]

\end{center}

\vspace{1mm}
\begin{abstract}
\noindent 
The two--dimensional quantum lattice Toda model
for the affine and simple Lie algebras of the
type A is considered. For its known L--operator
a correction of the second order in the lattice
parameter $\varepsilon$ is found. It is proved that 
the equation determining a correction of
the third order in $\varepsilon$ has no solutions.
\end{abstract}  

\vspace{5mm}
\section{Introduction}
\subsection{Continuous classical model}

The (1+1)--dimensional Toda chain associated with the
affine Lie algebra $A_{N-1}^{(1)}$ is a model that describes 
relativistic dynamics of $N$ scalar fields, $\phi_a$, $a=1,\ldots,N$, 
assigned to the nodes of the corresponding Dynkin diagram.
Their equations of motion are 
\begin{equation}\label{em1}
{} \bigl(\frac{\partial^{2}}{\partial t^{2}}-
 \frac{\partial^{2}}{\partial x^{2}} \bigr) \, \phi_a
    = \frac{2 m^2}{\beta} \, \bigl(
 e^{2\beta(\phi_{a+1} - \phi_a)} -
    e^{2\beta(\phi_a - \phi_{a-1} )} \bigr) \,.
\end{equation}
Here and below the index which enumerates the nodes of the 
affine Dynkin diagram takes values in ${\mathbb Z}/N$. 
In particular, we have $\phi_{N+a} \equiv \phi_{a}$.
Equations of motion (\ref{em1}) are generated by the 
following Hamiltonian and Poisson structure: 
\begin{align}
\label{ham1}
 & H = \sum\limits_{a=1}^{N} {\int} dx \Bigl(
    \fr{1}{2}\pi_a^2 + \fr{1}{2}\bigl(\partial_x\phi_a\bigr)^2
 + \fr{m^2}{\beta^2} \, e^{2\beta(\phi_{a+1} - \phi_a)} \Bigr) \,,  \\
\label{ps1}
 & \qquad\qquad \{\pi_a(x) , \phi_b(y)\} 
 = \delta_{ab} \, \delta(x-y) \,.
\end{align}

\vbox{The model under consideration is integrable. It admits 
the zero curvature representation with the following
$U$--$V$ pair~\cite{Mik,LS}
\begin{align}
\label{u1}
 U(\la) =&\   \sum_{a=1}^{N}
  \beta \, \pi_a \, e_{aa} +
 m \sum_{a=1}^{N} e^{\beta(\phi_{a+1} - \phi_a) }
 \bigl( \la^{\delta_{a,N}} e_{a,a+1} +
    \la^{-\delta_{a,N}} e_{a+1,a} \bigr) \,,\\ 
\label{v1}
 V(\la) =&\   \sum_{a=1}^{N}
  \beta \, \partial_x\phi_a \, e_{aa} +
  m \sum_{a=1}^{N} e^{\beta(\phi_{a+1} - \phi_a) }
 \bigl( \la^{\delta_{a,N}} e_{a,a+1} -
 \la^{-\delta_{a,N}} e_{a+1,a} \bigr) \,, 
\end{align} 
where $e_{ab}$ stands for the basis matrix such that
$(e_{ab})_{ij} = \delta_{ai}\delta_{bj}$.}

The matrix $U$ satisfies the following relation
(the so--called fundamental Poisson brackets, see \cite{FT}): 
\begin{equation}\label{fb1}
 \bigl\{ U_1(\la)   ,   U_2(\mu) \bigr\} =
   \bigl[ r \bigl( \fr{\la}{\mu} \bigr) ,
    U_1(\la)   +   U_2(\mu)\bigr] \,,
\end{equation}
where $r(\la)$ is the classical trigonometric r--matrix
for the algebra $A_{N-1}$, see~\cite{BVV,J1,FT}.
Here and below the lower indices denote the tensor
component, e.g. $U_1 = U \otimes \im$.

\subsection{Quantum lattice model}

The direct quantization of a continuous interacting 
field theory is known to have problems with 
ultraviolet divergences. 
A possible roundabout is to consider a discrete 
regularization of the model by putting it on 
the one dimensional lattice of a step~$\Delta$.
For the lattice model, the quantum canonical variables 
that sit at different sites commute, and those 
that sit at the same site
satisfy the following relations 
\begin{equation}\label{cr1}
 [\pi_a , \phi_b] = -i \, \hbar \, \delta_{ab} \,.
\end{equation}
The classical continuous limit of these relations recovers
the Poisson structure (\ref{ps1}) if one assumes that 
\begin{equation}\label{pfn}
  \pi_a^{(n)} = \Delta\, \pi_a(x) \,\qquad
  \phi_a^{(n)} = \phi_a(x) \,, \qquad x= n\Delta \,,
\end{equation}
where $n$ is the lattice site's number  
(it will be omitted in the subsequent formulae).

Given an integrable classical continuous model, 
its quantum lattice analogue is integrable as well
if there exist a quantum L--operator (see, e.g.
\cite{BIK}) such that:\\
i) its classical continuous limit recovers the 
corresponding matrix $U$:
\begin{equation}\label{lrn}
 L(\la)\Bigm|_{\hbar=0}
 =\im + \Delta \, U(\la) + o(\Delta)  \,;
\end{equation}
ii) it satisfies the following quadratic commutation 
relation which is a lattice analogue of the 
fundamental Poisson brackets (\ref{fb1}):
\begin{equation}\label{RLL}
  R(\fr{\la}{\mu}) \ L_1(\la) \, L_2(\mu) =
  L_2(\mu) \, L_1(\la) \ R(\fr{\la}{\mu}) \,.
\end{equation}

The quantum R--matrix must satisfy the Yang--Baxter relation, 
\begin{equation}\label{YB}
  R_{12}(\fr{\la}{\mu}) \, R_{13}(\la) \, R_{23}(\mu) =
  R_{23}(\mu) \, R_{13}(\la)\, R_{12}(\fr{\la}{\mu}) \,,
\end{equation}
and its classical limit must recover the classical r--matrix.

For the $A_{N-1}^{(1)}$ Toda model, the quantum R--matrix
is given by~\cite{BVV,J1}
\begin{equation}\label{Rmat}
{}  R(\la)
% & \equiv &
 %\sum_{a,b,c,d} R^{ab}_{cd}(\la) \, e_{ab} \ot e_{cd}
 = \sum_{a,b=1}^{N} 
(\la q^{\delta_{ab }}-q^{-\delta_{ab }} )\, e_{aa} \ot e_{bb}
+ (q-q^{-1})\, \sum_{a \neq b}^{N} \la^{\theta_{ab}} \,
    e_{ab} \ot e_{ba}  \,,
\end{equation}
where $q=e^{i \beta^2 \hbar}$,
and $\theta_{ab} = 0$ for $a<b$, $\theta_{ab} = 1$ for $a>b$.

\section{Lattice quantum L--operator}
\subsection{First order}

We will use the following notations:
\begin{align}\nn
{}& \Pi  =\text{diag}(\pi_1,\ldots,\pi_{N}) \,, &&
 \Phi =\text{diag}(\phi_1,\ldots,\phi_{N}) \,, \\[1mm]
 \nn 
{}& \hat{e}_a = \la^{\delta_{a,N}} e_{a,a+1} \,, &&
 \hat{f}_a = \la^{-\delta_{a,N}} e_{a+1,a} \,, \\ 
\nn 
{}& \hat{E} =\sum_{a=1}^N \hat{e}_a \,, &&
\hat{F} =\sum_{a=1}^N \hat{f}_a \,.
\end{align}

In the seminal work~\cite{J1} M.~Jimbo has found
an approximate quantum L--operator for 
the $A_{N-1}^{(1)}$ Toda model.
Namely, he showed that the following L--operator 
\begin{equation}\label{LJ}
\begin{aligned}
 L^{J}(\la) &
 %% &\equiv L^{(0)}(\la) + \ve\, L^{(1)}(\la) 
 = e^{\frac{\beta}{2} \Pi } \,
 \Bigl(\im +\ve \bigl(e^{-\beta{\rm ad}_{\Phi}}\hat{E}+
e^{\beta{\rm ad}_{\Phi}}\hat{F}\bigr) \Bigr)
  \, e^{\frac{\beta}{2} \Pi }  \\ 
{} &= \sum_{a=1}^{N} e_{aa} e^{\beta \pi_a } +
 \ve \, e^{\frac{\beta}{2} \Pi } \, \Bigl( \sum_{a=1}^{N}
     e^{\beta(\phi_{a+1} - \phi_a) }
 \bigl( \hat{e}_a + \hat{f}_a  \bigr)
 \Bigl) \, e^{\frac{\beta}{2} \Pi } 
\end{aligned}
\end{equation}
satisfies the RLL--relations (\ref{RLL}) in the
zeroth and first orders in~$\ve$.

It is easy to see that (\ref{LJ}) satisfies 
the condition (\ref{lrn}) if we set 
\begin{equation}\label{emd}
\ve = m \Delta
\end{equation}
and take into account 
the ``renormalization'' of momenta~(\ref{pfn})
in the continuous limit. 

Note that, although the L--operator (\ref{LJ}) is approximate,
the corresponding R--matrix (\ref{Rmat}) contains no small 
parameter $\ve$ and is an exact solution to~(\ref{YB}).
In order to treat the quantum Toda model by means of the
quantum inverse scattering method (see~\cite{BIK}) one needs
an exact quantum L--operator which solves the relation (\ref{RLL})
in all orders in~$\ve$.  In the present paper we will 
consider second and third order corrections 
to the L--operator~(\ref{LJ}). 

\subsection{Second order}

Consider an L--operator $L(\lambda,\ve)$ that
admits a series expansion in the parameter $\ve$,  
\begin{equation}\label{Leps}
 L(\lambda,\ve)= \sum_{n \geq 0} \ve^n L^{(n)}(\la) \,.
\end{equation}
Expanding relation (\ref{RLL}) in $\ve$, we
obtain an infinite set of relations for $L^{(n)}(\la)$. 
The explicit
form of those corresponding to the order
$\ve^n$, $n=0,1,2,3$ is
\begin{eqnarray}
\label{eps0}
 & 
  R(\fr{\lambda}{\mu}) \, 
   L^{(0)}_1(\lambda) \, L^{(0)}_2(\mu) =
   L^{(0)}_2(\mu) \, L^{(0)}_1(\lambda) 
   \, R(\fr{\lambda}{\mu})  \,, & \\[1mm]
\label{eps1}
& \begin{aligned}
{}
  R(\fr{\lambda}{\mu}) \, \bigl(
  L^{(1)}_1(\lambda) \, L^{(0)}_2(\mu) &+
   L^{(0)}_1(\lambda) \, L^{(1)}_2(\mu) \bigr) = \\
{}&
  \bigl( L^{(1)}_2(\mu) \, L^{(0)}_1(\lambda) 
   + L^{(0)}_2(\mu) \, L^{(1)}_1(\lambda) 
   \bigr) \, R(\fr{\lambda}{\mu})  \,, 
\end{aligned} &  \\[1mm]
& \begin{aligned}
{}
  R(\fr{\lambda}{\mu}) \, \bigl(
  L^{(2)}_1(\lambda) & \, L^{(0)}_2(\mu) +
  L^{(0)}_1(\lambda) \, L^{(2)}_2(\mu) +
   L^{(1)}_1(\lambda) \, L^{(1)}_2(\mu) \bigr) = \\
\label{eps2}
{}&
  \bigl( L^{(2)}_2(\mu) \, L^{(0)}_1(\lambda) 
   + L^{(0)}_2(\mu) \, L^{(2)}_1(\lambda) 
   + L^{(1)}_2(\mu) \, L^{(1)}_1(\lambda) 
   \bigr) \, R(\fr{\lambda}{\mu})  \,.
\end{aligned} & \\[1mm]
& \begin{aligned}
{}
  R(\fr{\lambda}{\mu}) &\, \bigl(
  L^{(3)}_1(\lambda)  \, L^{(0)}_2(\mu) +
  L^{(0)}_1(\lambda) \, L^{(3)}_2(\mu) +
   L^{(2)}_1(\lambda) \, L^{(1)}_2(\mu) 
   + L^{(1)}_1(\lambda) \, L^{(2)}_2(\mu)\bigr) = \\
\label{eps3}
{}&
  \bigl( L^{(3)}_2(\mu) \, L^{(0)}_1(\lambda) 
   + L^{(0)}_2(\mu) \, L^{(3)}_1(\lambda) 
   + L^{(2)}_2(\mu) \, L^{(1)}_1(\lambda) 
   + L^{(1)}_2(\mu) \, L^{(2)}_1(\lambda) 
   \bigr) \, R(\fr{\lambda}{\mu})  \,.
\end{aligned} &  
\end{eqnarray}

We will take
\begin{equation}\label{L01}
 L^{(0)}(\lambda) = e^{\beta\Pi} \,, \qquad
 L^{(1)}(\lambda) =
  e^{\frac{\beta}{2} \Pi } \,
 \bigl( \rho_+ \, e^{-\beta{\rm ad}_{\Phi}}\hat{E}+
 \rho_- \, e^{\beta{\rm ad}_{\Phi}}\hat{F}\bigr) 
  \, e^{\frac{\beta}{2} \Pi } \,.
\end{equation}
Notice that we slightly generalized the first 
order L--operator (\ref{LJ})
by introducing arbitrary coefficients 
$\rho_+$,~$\rho_-$. 
In order to comply with the classical limit
condition (\ref{lrn}) we have to assume that
$\rho_+, \rho_- \to 1$ as $\hbar \to 0$.

The problem which we want to solve is the following.
First, given $L^{(0)}(\lambda)$ and $L^{(1)}(\lambda)$
as in (\ref{L01}), find the most general solution 
$L^{(2)}(\lambda)$ to the equation~(\ref{eps2}). 
Then investigate whether, for some suitable 
$L^{(2)}(\lambda)$, equation~(\ref{eps3}) has a 
solution~$L^{(3)}(\lambda)$. 

The main result of the present article is the following
statement:\\[2mm]
{\bf Proposition 1}.\ {\em
Let $R(\la)$ be given by (\ref{Rmat}), and 
$L^{(0)}(\lambda)$, $L^{(1)}(\lambda)$ by~(\ref{L01}). 
Then \\
{\em\bf i)}\ The general solution to equation~(\ref{eps2})
is given by
\begin{equation}\label{LL} 
 \tilde{L}^{(2)}(\lambda) = 
 L^{(2)}(\lambda) + \tilde{L}^{(1)}(\lambda) \,.
\end{equation}
Here  $\tilde{L}^{(1)}(\lambda)$ 
is an arbitrary solution to 
equation~(\ref{eps1}), and $L^{(2)}(\lambda)$ is given by 
\begin{equation}
\begin{aligned} \label{L2ef}
 L^{(2)}(\la) =\ &
 e^{\frac{\beta}{2}\Pi} \Bigl( \gamma_1\, 
  e^{-\beta{\rm ad}_{\Phi}}\hat{E}^2
+ \gamma_2\, e^{\beta{\rm ad}_{\Phi}}\hat{F}^2 \\
{}& + \gamma_3\, 
(e^{-\beta{\rm ad}_{\Phi}}\hat{E})
 (e^{\beta{\rm ad}_{\Phi}}\hat{F})
 + \gamma_4\,  (e^{\beta{\rm ad}_{\Phi}}\hat{F})
  (e^{-\beta{\rm ad}_{\Phi}}\hat{E})
 \Bigr) e^{\frac{\beta}{2}\Pi}  \,,
\end{aligned}
\end{equation}
where the coefficients $\gamma_i$ must satisfy
the following conditions
\begin{align}\label{gg1}
 {}& \text{for}\ N=2: \qquad 
  \gamma_3 + \gamma_4 = \rho_+  \rho_-  \,, \\
\label{gg2}
 {}& \text{for}\ N \geq 3: \qquad
  \gamma_1 = \frac{q \, \rho_+^2}{1+q} \,, \qquad 
  \gamma_2 = \frac{\rho_-^2}{1+q} \,, \qquad
  \gamma_3 + \gamma_4 = \rho_+  \rho_- \,. 
\end{align}
{\em\bf ii)}\ For any choice of 
$\tilde{L}^{(1)}(\lambda)$ in (\ref{LL}), equation
(\ref{eps3}) has no solution for $L^{(3)}(\lambda)$.
} \\[2mm]
\indent 
Proof is given in Appendix~A.

Formula (\ref{LL}) reflects the fact that the general
solution to an inhomogeneous equation is the sum
of its particular solution and the general
solution of the corresponding homogeneous equation.
Let us remark that 
$\tilde{L}^{(1)}(\lambda)$ does not have to
satisfy the condition~(\ref{lrn}). 

The explicit expression for (\ref{L2ef}) involving 
the basis matrices is
\begin{align}
\label{L2aa}
 L^{(2)}(\la)  =\ {}&
 \sum_{a=1}^{N} 
   e^{\frac{\beta}{2}\pi_a} \bigl(
   \gamma_3 \, e^{2\beta(\phi_{a+1}-\phi_a)} + 
   \gamma_4\, e^{2\beta(\phi_{a}-\phi_{a-1})} \bigr)
   e^{\frac{\beta}{2}\pi_a}   \, e_{aa} \\
 \nonumber
{}& + \gamma_1 \sum_{a=1}^{N} 
 e^{\frac{\beta}{2}\pi_{a-1}}
e^{\beta(\phi_{a+1}-\phi_{a-1})}e^{\frac{\beta}{2}\pi_{a+1}}
 \, e_{a-1,a+1} \, \lambda^{\delta_{a,1}+\delta_{a,N}}  \\
\nonumber 
{}& + \gamma_2 \sum_{a=1}^{N} 
 e^{\frac{\beta}{2}\pi_{a+1}}
e^{\beta(\phi_{a+1}-\phi_{a-1})}e^{\frac{\beta}{2}\pi_{a-1}}
 \, e_{a+1,a-1} \, \lambda^{\delta_{a,1}+\delta_{a,N}}  \,.
\end{align} 

For $N=2$, eq. (\ref{L2aa}) contains only diagonal terms.
In this case, choosing $\rho_+=\rho_-=1$,
$\gamma_1=\gamma_2=0$, and $\gamma_3=1$
(or $\gamma_3=0$), we obtain  an exact L--operator, 
\begin{equation}\label{Lsh}
 L(\lambda)= \left( \begin{array}{cc}
 e^{\frac{\beta}{2}\pi_1} \bigl( 1+
 \ve^2 e^{2\tilde{\beta}(\phi_{2} -\phi_{1})}\bigr) 
 e^{\frac{\beta}{2}\pi_1} &
 \ve e^{\frac{\beta}{2}\pi_1} \bigl(
 e^{\beta(\phi_{2} -\phi_{1})} +
 \la^{-1} e^{\beta(\phi_{1} -\phi_{2})}
 \bigr) e^{\frac{\beta}{2}\pi_2} \\
 \ve e^{\frac{\beta}{2}\pi_2} \bigl(
 e^{\beta(\phi_{2} -\phi_{1})}+
 \la e^{\beta(\phi_{1} -\phi_{2})}
 \bigr) e^{\frac{\beta}{2}\pi_1}
 & e^{\frac{\beta}{2}\pi_2} \bigl( 1+
 \ve^2 e^{2\tilde{\beta}(\phi_{1} -\phi_{2})}\bigr) 
 e^{\frac{\beta}{2}\pi_2}
 \end{array}\right) \,,
\end{equation}
where $\tilde{\beta} =(\gamma_3-\gamma_4)\beta$. 
L--operator (\ref{Lsh}) satisfies relation (\ref{RLL}) 
in all orders in~$\ve$.
Upon the reduction $\phi_2=-\phi_1$, $\pi_2=-\pi_1$, 
eq. (\ref{Lsh}) yields the well--known exact L--operator 
for the sinh--Gordon model~\cite{Kul_Resh,BIK}.

\section{Reduction to non--affine case}
The (1+1)--dimensional Toda chain associated with the
simple Lie algebra $A_{N-1}$ describes 
relativistic dynamics of $N$ scalar fields  
whose equations of motion are given by the same 
equation (\ref{em1}) where no periodicity in the
index $a$ is assumed. 
In this case one can formally set 
$\beta\phi_0 = -\beta\phi_{N+1} = +\infty$
in (\ref{em1}) and~(\ref{ham1}).
The same procedure applied to (\ref{u1})--(\ref{v1}) 
yields the U--V pair without a spectral parameter. 

In order to keep the spectral parameter in the
U--V pair, the following procedure was suggested
in \cite{FTi} (in the case corresponding to $A_{1}$; 
a generalization
was considered in~\cite{AKS}). Take $\xi>0$  
and shift (the zero modes of) the fields,
the mass and the spectral parameter 
in (\ref{u1})--(\ref{v1}) as follows,
\begin{equation}\label{phixi} 
  \phi_a \to \phi_a + a\, \xi/\beta \,, \qquad
  m \to e^{-\xi} \, m \,, \qquad
  \la \to e^{\xi N}  \la \,.
\end{equation}
Then the limit $\xi \to +\infty$ yields
the following U--V pair
\begin{equation}\label{UVn} 
 U(\la) = \beta\Pi + m 
 \bigl(e^{-\beta{\rm ad}_{\Phi}}\hat{E}+
e^{\beta{\rm ad}_{\Phi}} {F}\bigr) , \qquad  
V(\la) = \beta\partial_x \Phi + m 
 \bigl(e^{-\beta{\rm ad}_{\Phi}}\hat{E} -
e^{\beta{\rm ad}_{\Phi}} {F}\bigr) , 
\end{equation}
where  
\begin{equation}\label{EFn} 
\hat{E} =\sum_{a=1}^N \hat{e}_a 
 = \sum_{a=1}^N  \la^{\delta_{a,N}} e_{a,a+1} \,, \qquad
 F =\sum_{a=1}^{N-1} \hat{f}_a =
 \sum_{a=1}^{N-1} e_{a+1,a} .
\end{equation} 

The U--matrix in (\ref{UVn}) satisfies the
same fundamental Poisson bracket (\ref{fb1})
with the same classical r--matrix as in the affine case.
%This is immediate from the following statement. 

In the non--affine case we have the following counterpart
of Proposition~1.\\[2mm]
{\bf Proposition 2}.\ {\em
Let $R(\la)$ be given by (\ref{Rmat}). Then \\
\em{\bf i)}\
Equations (\ref{eps0}) and (\ref{eps1}) 
admit the following solutions
\begin{equation}\label{L01n}
 L^{(0)}(\lambda) = e^{\beta\Pi} \,, \qquad
 L^{(1)}(\lambda) =
  e^{\frac{\beta}{2} \Pi } \,
 \bigl( \rho_+ \, e^{-\beta{\rm ad}_{\Phi}}\hat{E}+
 \rho_- \, e^{\beta{\rm ad}_{\Phi}} F \bigr) 
  \, e^{\frac{\beta}{2} \Pi } \,.
\end{equation}
\em{\bf ii)}\ Given 
$L^{(0)}(\lambda)$, $L^{(1)}(\lambda)$ as in (\ref{L01n}), 
the general solution to equation~(\ref{eps2})
is given by
\begin{equation}\label{LLn} 
 \tilde{L}^{(2)}(\lambda) = 
 L^{(2)}(\lambda) + \tilde{L}^{(1)}(\lambda) \,.
\end{equation}
Here  $\tilde{L}^{(1)}(\lambda)$ 
is an arbitrary solution to 
equation~(\ref{eps1}), and $L^{(2)}(\lambda)$ is given by 
\begin{equation}
\begin{aligned} \label{L2efn}
 L^{(2)}(\la) =\ &
 e^{\frac{\beta}{2}\Pi} \Bigl( \gamma_1\, 
  e^{-\beta{\rm ad}_{\Phi}}\hat{E}^2
+ \gamma_2\, e^{\beta{\rm ad}_{\Phi}} F^2 \\
{}& + \gamma_3\, 
(e^{-\beta{\rm ad}_{\Phi}}\hat{E})
 (e^{\beta{\rm ad}_{\Phi}} F )
 + \gamma_4\,  (e^{\beta{\rm ad}_{\Phi}} F)
  (e^{-\beta{\rm ad}_{\Phi}}\hat{E})
 \Bigr) e^{\frac{\beta}{2}\Pi}  \,,
\end{aligned}
\end{equation}
where the coefficients $\gamma_i$ must satisfy
conditions (\ref{gg1}) and~(\ref{gg2}). \\[1mm]
{\em\bf iii)}\ For any choice of 
$\tilde{L}^{(1)}(\lambda)$ in (\ref{LLn}), equation
(\ref{eps3}) has no solution for $L^{(3)}(\lambda)$.
} \\[2mm]
\indent 
Proof is given in Appendix.

For $N=2$, eq. (\ref{L2efn}) contains only diagonal terms.
Furthermore, $F^2=0$.
In this case, choosing $\rho_+=\rho_-=1$,
$\gamma_1=0$, and $\gamma_3=1$
(or $\gamma_3=0$), we obtain  an exact L--operator, 
\begin{equation}\label{Lli}
 L(\lambda)= \left( \begin{array}{cc}
 e^{\frac{\beta}{2}\pi_1} \bigl( 1+
 \ve^2 e^{2\tilde{\beta}(\phi_{2} -\phi_{1})}\bigr) 
 e^{\frac{\beta}{2}\pi_1} &
 \ve e^{\frac{\beta}{2}\pi_1}  
 e^{\beta(\phi_{2} -\phi_{1})}  
   e^{\frac{\beta}{2}\pi_2} \\
 \ve e^{\frac{\beta}{2}\pi_2} \bigl(
 e^{\beta(\phi_{2} -\phi_{1})}+
 \la e^{\beta(\phi_{1} -\phi_{2})}
 \bigr) e^{\frac{\beta}{2}\pi_1}
 & e^{\frac{\beta}{2}\pi_2} \bigl( 1+
 \ve^2 e^{2\tilde{\beta}(\phi_{1} -\phi_{2})}\bigr) 
 e^{\frac{\beta}{2}\pi_2}
 \end{array}\right) \,,
\end{equation}
where $\tilde{\beta} =(\gamma_3-\gamma_4)\beta$. 
L--operator (\ref{Lli}) satisfies relation (\ref{RLL}) 
in all orders in~$\ve$.
Upon the reduction $\phi_2=-\phi_1$, $\pi_2=-\pi_1$, 
eq. (\ref{Lli})
yields the exact L--operator for the Liouville
model~\cite{FTi}.

\par\vspace*{4mm}\noindent
{\small
{\bf Acknowledgements.}
\noindent 
This work has been supported in part by the 
Russian Foundation for Fundamental Research 
(grants 09-01-93108,
11-01-00570, and 11-01-12037), and by 
the Swiss National Science Foundation
(grants 200020-126909 and PDFMP2-137071).
}

%%%%%%%%%%%%%%%%%%%%%%%%%%%%%%%%%
\appendix
\section{Appendix }
\subsection{Proof of the Proposition 1. Second order} 
We will use the following notations:
\begin{align}\nn
{}& H_a  = e_{aa} - e_{a+1,a+1} \,, \qquad
 K_a = q^{\frac{1}{2} H_a} \,, \qquad
 \alpha_a(X) = \tr (H_a X)  \,,
\end{align}
where $a=1,\,{\ldots}\,,N$ and 
$e_{N+1,N+1} \equiv e_{11}$.
Then we have
\begin{equation}\label{efK} 
 K_a \hat{e}_b =  q^{\frac{1}{2} A_{ab}} \,
  \hat{e}_b  K_a \,, \qquad
 K_a \hat{f}_b =  q^{-\frac{1}{2} A_{ab}} \,
  \hat{f}_b  K_a \,,  
\end{equation}
where $A$ is the Cartan matrix of the 
affine algebra~$A_{N-1}^{(1)}$.

As the first step, following \cite{J1}, 
we rewrite $L^{(1)}(\lambda)$
and $L^{(2)}(\lambda)$
by moving $e^{\frac{\beta}{2}\Pi}$ to the
extreme right,
\begin{align}\label{LJ2b}
 L^{(1)}(\lambda) &=  
  \sum_{a=1}^N  e^{-\beta \alpha_a(\Phi)}
   \Bigl( \rho_+ 
   e^{\frac{\beta}{2} \alpha_a(\Pi)}  K_a \hat{e}_a 
   + \rho_- 
   e^{-\frac{\beta}{2}  \alpha_a(\Pi)}  K_a \hat{f}_a \Bigr)
   e^{\beta \Pi} , \\
\label{LJ2c}
L^{(2)}(\lambda) &= \sum_{a=1}^N \Bigl(
  \gamma_1\, e^{-\beta (\alpha_a(\Phi)+\alpha_{a+1}(\Phi))}
   e^{\frac{\beta}{2} (\alpha_a(\Pi)+\alpha_{a+1}(\Pi))}
  K_a K_{a+1} \hat{e}_a \hat{e}_{a+1} \\
\nn
{}& \qquad\quad +
  \gamma_2\, e^{-\beta (\alpha_a(\Phi)+\alpha_{a+1}(\Phi))}
   e^{-\frac{\beta}{2} (\alpha_a(\Pi)+\alpha_{a+1}(\Pi))}
    K_a K_{a+1} \hat{f}_{a+1} \hat{f}_a \\[1mm]
    \nn
{}& \qquad\quad +  e^{-2\beta \alpha_a(\Phi)}
 K^2_a \bigl( \gamma_3\, \hat{e}_{a} \hat{f}_{a} 
  + \gamma_4\, \hat{f}_{a} \hat{e}_{a} \bigr)
   \Bigr)  e^{\beta \Pi} .
\end{align} 
Next we substitute (\ref{LJ2b})--(\ref{LJ2c}) 
into (\ref{eps1})--(\ref{eps2}) and move all the 
factors containing 
$e^{\beta \Pi}$  to the right using the relations
\begin{equation}\nn
 e^{\beta \Pi_1}  e^{-\beta \alpha_a(\Phi)}
  = e^{-\beta \alpha_a(\Phi)} 
  \bigl(K_a^2 \otimes \im)\, e^{\beta \Pi_1} , \qquad
e^{\beta \Pi_2}  e^{-\beta \alpha_a(\Phi)}
  = e^{-\beta \alpha_a(\Phi)} 
  \bigl(\im \otimes K_a^2)\, e^{\beta \Pi_2} .
\end{equation}
Finally,  
matching the coefficients at functionally independent
exponentials of quantum fields, we obtain a set of
relations. Here one should take into account that $R(\lambda)$
commutes with $(e_{aa} \otimes \im + \im \otimes e_{aa})$
and hence
\begin{align}\nn
  [ R(\lambda), K_a \otimes K_a] = 0 \,, \qquad
  [ R(\lambda), e^{\beta \Pi_1} e^{\beta \Pi_2}] = 0 \,.
\end{align}
The relations that arise  
as matching conditions for the coefficients
in (\ref{eps1})
at the fields $e^{-\beta \alpha_a(\Phi)}
   e^{\pm \frac{\beta}{2} \alpha_a(\Pi)}$ are
\begin{equation}\label{Rdx}
   R(\fr{\lambda}{\mu}) \, \Delta(x_a)
   = \Delta^\prime(x_a) \, R(\fr{\lambda}{\mu}) \,,
    \qquad a = 1,\,{\ldots}\,,N \,,
\end{equation}
where
$x_a = \hat{e}_a, \hat{f}_a$, respectively, and
\begin{equation}\label{Def}
 \Delta(x_a) =   x_a \otimes K^{-1}_a + 
    K_a \otimes x_a  \,, \qquad
  \Delta^\prime(x_a) =  
  x_a \otimes K_a +  K_a^{-1} \otimes x_a  \,. 
\end{equation}
Here and below $x_N \otimes \im$ depends on $\lambda$
while $\im \otimes x_N$ depends on~$\mu$.  

In \cite{J1}, Jimbo has shown that the solution
to equations (\ref{Rdx}) is unique up to an overall scalar
factor and that it is given by the R--matrix~(\ref{Rmat}).

Now, treating  equation (\ref{eps2}) similarly and
matching the coefficients at the fields
\begin{equation}\nn
e^{-\beta (\alpha_a(\Phi)+\alpha_{b}(\Phi))}
   e^{\frac{\beta}{2} (\kappa_1 \alpha_a(\Pi)
    +\kappa_2 \alpha_{b}(\Pi))} , \qquad
    \kappa_i = \pm \,,
\end{equation}
we find the relations
\begin{equation}\label{RXY}
 R(\fr{\lambda}{\mu}) \, X_{ab}^{\kappa_1 \kappa_2} =
 \bigl(X_{ab}^{\kappa_1 \kappa_2}\bigr)^\prime  \, 
 R(\fr{\lambda}{\mu})  \,, 
 \qquad a,b = 1,\,{\ldots}\,,N \,,
\end{equation}
where the prime on the r.h.s. denotes 
the permutation
of the tensor factors (analogous to that in~(\ref{Def})).
Obviously, $X^{\kappa_1 \kappa_2}_{ab}=
 X^{\kappa_2 \kappa_1}_{ba}$.
We have
\begin{align}
\label{Xaa}
{}& X^{++}_{ab} =  
\hat{e}_a K_{b} \otimes K_a^{-1} \hat{e}_{b}
 + (1-\delta_{ab})\, 
  \hat{e}_{b} K_{a} \otimes K_{b}^{-1} \hat{e}_a 
  , \qquad \text{при}\ \
    a-b \neq \pm 1 \bmod N ,\\[1mm]
\label{Xe}
{}& X^{++}_{a,a+1} = 
   \gamma_1 \, \bigl(
  \hat{e}_a \hat{e}_{a+1} \otimes K_a^{-1} K_{a+1}^{-1}
 +   K_a K_{a+1} \otimes \hat{e}_a \hat{e}_{a+1}  
 \bigr) \\
\nn
{}& \qquad\qquad + \rho_+^2 \bigl(
 \hat{e}_a K_{a+1} \otimes K_a^{-1} \hat{e}_{a+1}
 + \hat{e}_{a+1} K_{a} \otimes K_{a+1}^{-1} \hat{e}_a 
 \bigr)   \,,\\[1mm]
  \label{Yaa}
{}& X^{--}_{ab} =   
 \hat{f}_a K_{b} \otimes K_a^{-1} \hat{f}_{b}
 + (1-\delta_{ab})\, 
 \hat{f}_{b} K_{a} \otimes K_{b}^{-1} \hat{f}_a  
 , \qquad \text{при}\ \ 
  a-b \neq \pm 1 \bmod N  , \\[1mm]
 \label{Yf}
{}&  X^{--}_{a,a+1} = 
 \gamma_2\, \bigl( 
  \hat{f}_{a+1} \hat{f}_a
  \otimes K_a^{-1} K_{a+1}^{-1} + 
  K_a K_{a+1} \otimes \hat{f}_{a+1} \hat{f}_a
  \bigr)  \\
\nn
{}& \qquad\qquad  + \rho_-^2 \bigl( 
 \hat{f}_a K_{a+1} \otimes K_a^{-1} \hat{f}_{a+1}
+ \hat{f}_{a+1} K_{a} \otimes K_{a+1}^{-1} \hat{f}_a 
 \bigr)  \,, \\[1mm]
\label{Zef}
{}&  X^{+-}_{ab} = 
 \delta_{ab} \Bigl(
 \gamma_3 \,\bigl( \hat{e}_a \hat{f}_a \otimes K_a^{-2}
  +  K_a^{2} \otimes \hat{e}_a \hat{f}_a \bigr)
+  \gamma_4 \,\bigl( \hat{f}_a \hat{e}_a \otimes K_a^{-2}
  +  K_a^{2} \otimes \hat{f}_a \hat{e}_a \bigr) \Bigr) \\
\nn  
{}& \qquad\qquad  + \rho_+ \rho_- 
 \bigl( \hat{e}_a K_b \otimes K_a^{-1} \hat{f}_b
+ \hat{f}_b K_a \otimes K_b^{-1} \hat{e}_a \bigr) \,.
\end{align} 

Let us remark that eqs. (\ref{efK})
and (\ref{Def}) are relations for the generators of 
the affine algebra $A_{N-1}^{(1)}$ that
hold for any rank and representation.
However, in the case of the fundamental 
representation we have extra relations:
\begin{align*}
{}& \text{for}\ N \geq 2: \qquad  
\hat{e}_{a} \hat{f}_{b} = 
\hat{f}_{b}  \hat{e}_{a} =0  \,, 
\qquad \text{if}\ \  b \neq a \,,\\
{}& \text{for}\ N \geq 3: \qquad 
\hat{e}_{a} \hat{e}_{b} = 
\hat{f}_{b} \hat{f}_{a} =0 \,, 
\qquad \text{if}\ \  b-a \neq 1  \bmod N \,.
\end{align*}
Taking them into account, we observe that
\begin{align}
\label{Xaaa}
{}&  (1+q^2) X^{++}_{aa} =
 \Delta(\hat{e}_a) \Delta( \hat{e}_{a}) \,, \qquad
 (1+q^2) X^{--}_{aa} =
 \Delta(\hat{f}_a) \Delta( \hat{f}_{a}) \,, \\
{}& X^{++}_{ab} = \Delta(\hat{e}_a) 
 \Delta( \hat{e}_{b}) \,, \qquad
 X^{--}_{ab} = \Delta(\hat{f}_a) 
 \Delta( \hat{f}_{b}) \,, \qquad \text{при}\ \
 a-b \neq 0,\pm 1 \bmod N  \,, \\
 \label{Zaa}
{}&  X^{+-}_{ab} = \rho_+ \rho_- \, 
\Delta(\hat{e}_a)  \Delta( \hat{f}_{b}) \,, 
 \qquad \text{при}\ \   a \neq b \,,
\end{align}
and
\begin{align}
 \label{Xee} {}&
\begin{aligned}
{}& X^{++}_{a,a+1} -
 \gamma_1\, \Delta(\hat{e}_a) \Delta( \hat{e}_{a+1})
 -q(\rho_+^2-\gamma_1)\,  
 \Delta(\hat{e}_{a+1}) \Delta( \hat{e}_{a}) \\
{}& \qquad\qquad\qquad
 = \bigl((1-q) \rho_+^2 + (q-q^{-1})\gamma_1 \bigr)\, 
 \hat{e}_{a+1} K_{a} \otimes K_{a+1}^{-1} \hat{e}_a \,,
\end{aligned}  \\[1mm] 
 \label{Yff} {}&
\begin{aligned}
{}& X^{--}_{a,a+1} -
 \gamma_2\, \Delta(\hat{f}_{a+1}) \Delta( \hat{f}_a)
 - q^{-1}(\rho_-^2 -\gamma_2)\, 
 \Delta(\hat{f}_a) \Delta( \hat{f}_{a+1}) \\
{}& \qquad\qquad\qquad
 = \bigl( (1-q^{-1})\rho_-^2 + 
 (q^{-1}-q)\gamma_2 \bigr)\, 
 \hat{f}_{a} K_{a+1} \otimes K_{a}^{-1} \hat{f}_{a+1} \,,
\end{aligned}  \\[1mm]
 \label{Zeeff}  {}&
\begin{aligned}
{}& X^{+-}_{aa} -
 \gamma_3 \, \Delta(\hat{e}_a) \Delta(\hat{f}_a)
 - \gamma_4 \, \Delta(\hat{f}_a) \Delta(\hat{e}_a) \\
{}&  \qquad\qquad\qquad
 = (1-\gamma_3-\gamma_4) \, 
 \bigl (\hat{e}_a K_a \otimes K_a^{-1} \hat{f}_a +
 \hat{f}_a K_a \otimes K_a^{-1} \hat{e}_a \bigr)  \,.
 \end{aligned}
\end{align}

Thus the condition that $L^{(2)}$ under consideration 
is a solution to (\ref{eps2})
is equivalent to the requirement that relation (\ref{RXY})
holds for the r.h.s. of (\ref{Xaaa})--(\ref{Zeeff}).
Equations (\ref{Rdx})--(\ref{Def}) imply that the r.h.s. of (\ref{Xaaa})--(\ref{Zaa}) does satisfy~(\ref{RXY}). 
Furthermore,  
 it is straightforward to check that (\ref{RXY})
does not hold for the r.h.s. of (\ref{Xee})--(\ref{Zeeff}).
This implies that the scalar factors on the r.h.s. of 
these equations must vanish. Whence we obtain the
values of $\gamma_i$ given in the Proposition~1. 
  
\subsection{Proof of the Proposition 1. Third order}  
{\bf Lemma 1}.\ {\em 
Let $R(\la)$ be given by (\ref{Rmat}), and 
$L^{(0)}(\lambda)$ be as in~(\ref{L01}). 
Let 
\begin{equation}\label{Lee}
 \tilde{L}^{(1)}(\lambda) = \sum_{a,b=1}^N
  \tilde{L}^{(1)}_{ab}(\lambda) \, e_{ab}
\end{equation}
be  
an arbitrary solution to equation~(\ref{eps1}).
Then the operator--valued coefficients 
$\tilde{L}^{(1)}_{ab}(\lambda)$ 
vanish unless $a=b$ or $a-b = \pm 1 \bmod N$.}\\[2mm]
{\bf Proof.}\
Consider the matrix entry $e_{cb} \otimes e_{ac}$
of equation~(\ref{eps1}). Choose such $a,b,c$ that
$a \neq b$, $a \neq c$, $b \neq c$. 
Then,
since $L^{(0)}(\lambda)$ is a diagonal matrix, 
the computation
of the matrix element in question involves only
the non--diagonal part of the R--matrix~(\ref{Rmat}).
It is straightforward to check that as the result
we obtain the equation
\begin{equation}\label{L1ab} 
  \Bigl(\frac{\lambda}{\mu}\Bigr)^{\theta_{ca}} \,
  \tilde{L}^{(1)}_{ab}(\lambda) = 
  \Bigl(\frac{\lambda}{\mu}\Bigr)^{\theta_{cb}} \,
  \tilde{L}^{(1)}_{ab} (\mu) \,.
\end{equation}
Now, if $b-a \neq 0,\pm 1 \bmod N$, then (\ref{L1ab}) 
for $c=a\,{-}\,1 \bmod N$ and $c=a\,{+}\,1 \bmod N$
yields two equations that are inconsistent
unless $\tilde{L}^{(1)}_{ab}(\lambda)=0$.
This completes the proof of the Lemma. \qed \\[2mm]
\indent
In order to prove the part {\bf ii)} of the 
Proposition~1, we write
\begin{equation}\label{Lee2}
 \tilde{L}^{(2)}(\lambda) = \sum_{a,b=1}^N
  \tilde{L}^{(2)}_{ab}(\lambda) \, e_{ab} \,,\qquad
 L^{(3)}(\lambda) = \sum_{a,b=1}^N
  L^{(3)}_{ab}(\lambda) \, e_{ab} \,.
\end{equation}
for the general solution of (\ref{eps2}) and the
sought for solution of~(\ref{eps3}). 

Consider the matrix entry $e_{a,a+1} \otimes e_{a-1,a+1}$
of equation~(\ref{eps3}) in the $N \geq 3$ case. 
It is straightforward  to check 
that the resulting equation reads
\begin{equation}\label{Lff}
\begin{aligned} 
{}& (\frac{\lambda}{\mu}-1)\, 
L^{(1)}_{a,a+1}(\lambda) \tilde{L}^{(2)}_{a-1,a+1}(\mu) +
(q- q^{-1})\, \frac{\lambda}{\mu} \,
\tilde{L}^{(2)}_{a-1,a+1}(\lambda)  L^{(1)}_{a,a+1}(\mu) \\
{}& \qquad\qquad\qquad =
(q \frac{\lambda}{\mu} - q^{-1})\,  
\tilde{L}^{(2)}_{a-1,a+1}(\mu)  L^{(1)}_{a,a+1}(\lambda)\,.
\end{aligned} 
\end{equation}
Note that this equation, although coming from the
third order in the $\ve$--expansion, does not involve
matrix entries of $L^{(3)}$. The reason is that 
in (\ref{eps3})
$L^{(3)}$ is coupled to $L^{(0)}$ for which
the matrix entries $(a,a+1)$ and $(a-1,a+1)$ vanish. 

Now, by Lemma~1, we can replace $\tilde{L}^{(2)}$ 
in (\ref{Lff}) with the particular $L^{(2)}$ given 
by~(\ref{L2ef}) since they must have coinciding
matrix entries $(a-1,a+1)$.
Finally, it is easy to check that (\ref{Lff})
does not hold for the matrix entries of $L^{(1)}$ 
and $L^{(2)}$ (cf. (\ref{LJ}) and (\ref{L2aa})).
Therefore, for any possible choice of~$\tilde{L}^{(2)}$,
equation (\ref{eps3}) has no solution $L^{(3)}$.
\qed

\subsection{Proof of Proposition 2} 
 
Part {\bf i)}.\ 
The $L^{(1)}$ in (\ref{L01n}) can be 
obtained from $L^{(1)}$ in (\ref{L01}) by
setting $\hat{f}_{N} =0$.
Since relations (\ref{Rdx}) are linear in $x_a$,
they are consistent with such a reduction.
Hence it follows that $L^{(1)}$ given by (\ref{L01n})
is a solution to~(\ref{eps2}).

Part {\bf ii)}.\ 
Analogously, setting $\hat{f}_{N} =0$ 
in (\ref{L2ef}), we obtain~(\ref{L2efn}).
The direct inspection of (\ref{Xaa})--(\ref{Zef}) 
shows that $X^{++}_{ab}$ are not affected by
the reduction while $X^{--}_{ab}$ and $X^{+-}_{ab}$ 
vanish if $a=N$ or $b=N$
and do not change if $a,b \neq N$. 
Therefore relations (\ref{RXY}) remain
valid which, in turn, implies that (\ref{eps3}) holds.

Part {\bf iii)}.\ It suffices to repeat the
arguments given in Section~A.2 and notice that
the matrix entries 
$L^{(1)}_{a,a+1}$ and 
$L^{(2)}_{a-1,a+1}$ are not affected by
the reduction.   \qed

\end{document}